# Expansion of Graphene-Based Device Technology for Resistance Metrology

**Albert F. Rigosi**

Physical Measurement Laboratory, National Institute of Standards and Technology, Gaithersburg, MD 20899, United States of America

## 1. Assembly of *p-n* Junctions in Graphene-Based Devices

The field of Quantum Hall metrology had a strong start with the implemntation of GaAs-based devices, given that 2D materials systems provided access to interesting quantum phenomena [1-31], including the infrastructure associated with making relevant measurements. With the technology laid out, further improvements in both infrastructure and standards were achieved in the previous two decades as EG-based QHR devices became established as national standards [32-59]. Since the metrology community has reached some understanding that a comparison against GaAs-based QHR devices had been accomplished, the next steps became clearer as far as how the EG-based QHR with a single Hall bar could be further developed. Since the early 90s, it has been of modest interest that QHR devices have a means of interconnecting several single Hall bar elements and has since been a subject of research [60]. NMIs are now presently at a juncture where consideration must be granted beyond just simplicity of operation. A natural direction for resistance standards would be to increase the total accessible parameter space. This means using EG-based QHR devices to output more than the single value at the ν = 2 plateau (about 12.9 kΩ). A first natural question is whether one may use the ν = 6 plateau or ν = 10 plateau, and though some work has been done with these Landau levels in graphene, they simply do not offer the same level of precision as the ν = 2 plateau [32, 61].

One alternative for outputting new values is by using *p-n* junctions (*pn*Js). This species of device has been shown to possibly provide useful resistance quantization [62-66]. Woszczyna *et al*. approached this question a decade ago [63], noting that since metallic leads had to cross paths in traditional GaAs-based devices, typical fabrication methods would be overly complicated, having to include some form of multilayer interconnect technology. Any leakage from such technology would likely generate an additional Hall voltage contribution from the primary contact, thus becoming a detriment to the achievable uncertainty. Another practical concern for this devices interconnected with conventional metals would stem from the use of leads whose small resistances would eventually accumulate, potentially disrupting precision measurements. Graphene offered the opportunity to combine two regions of varying charge carrier polarity. Rather than requiring interconnections, such a device would only need a set of uniform top gates to modulate regions accordingly. This work demonstrated that *pn*Js were possible and worth exploring as a viable extension for QHR standards.

As demonstrated by Hu *et al*., EG-based devices can be fabricated on the order of 100 μm or less, rendering it possible to implement reliable top gates for adjusting the carrier density in various regions [62]. The underlying physics of these *pn*Js makes it possible to construct devices that can access quantized resistance values that are fractional or integer values of $R_K$. Hu *et al*. made one such assessment for a *pn*J device. The example data are shown in Fig. 1. A DCC was used to provide turn-key resistance traceability, and in this measurement, a four-terminal bridge configuration was used to compare against a 10 kΩ standard resistor (see Fig. 1 (a)). The measurement time for each data point with the DCC was 15 min, with an orange shaded region representing Fig. 1 (b), which clarifies the deviation of the DCC measurements with respect to zero. In Fig. 1 (a), the right axis is represented by black points and gives the relative uncertainty of each measurement as a function of source-drain current. For this *pn*J device, a precision of about $2 \times 10^7$ was achieved. And though this may be one or two orders of magnitude below what is possible for a conventional Hall bar device, one must recall that, based on this work, a programmable resistance standard may be built using the demonstrated techniques. Such flexibility and expansion of accessible parameter space could justify its further technical exploration within resistance metrology.

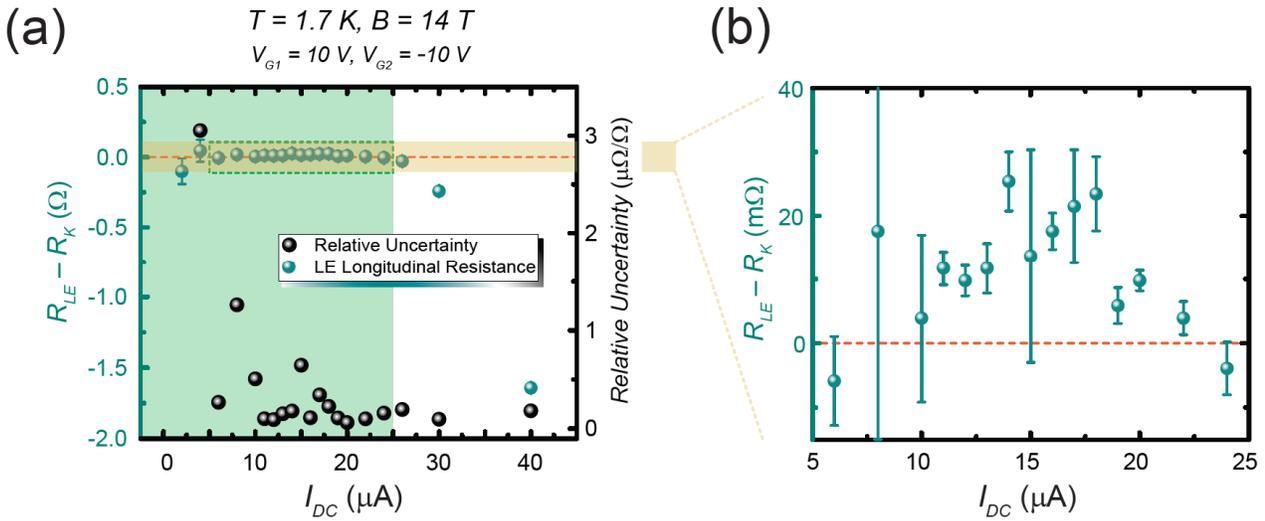

Figure 1. (a) The lower edge (LE) of the device in Hu *et al.* was compared against a 10 kΩ standard resistor with a DCC to give an assessment of the quality of quantization exhibited by the device. The resistor was selected based on its traceability to a quantum resistance standard at the National Institute of Standards and Technology. The turquoise points show the DCC measurements as deviations from $R_K$ on the left axis and the relative uncertainties of those deviations with DC current. The relative uncertainties improve with increasing current, but the device loses its optimal quantization after the critical current of 24 μA. The shaded green area indicates the well-quantized region. (b) The beige shaded area in (a) is magnified to show the deviations' error bars as well as the reference to zero deviation, marked as an orange dashed line. The error bars represent a 1σ deviation from the mean, where each data point represents an average of a set of data taken at each value of current. Ref. [62] is an open access article distributed under the terms of the Creative Commons CC BY license, which permits unrestricted use, distribution, and reproduction in any medium.

These types of devices are not as well-studied as other conventionally prepared devices, but given their interesting and rather unique properties, *pn*J devices offer a second, more fundamental path to avoid any resistance from interconnecting metallic contacts and multiple device connections. One recent approach by Momtaz *et al.* demonstrates how a programmable quantum Hall circuit could be designed to implement an iterative voltage bisection scheme, thus permitting the user to access any binary fraction of the ν = 2 plateau resistance to be obtained [66]. Their proposed circuit designs offer potential advantages for resistance metrology, as summarized here: first, their circuit contain no internal Ohmic contacts, a problem that has plagued metrologists for years. Thankfully, this design utilizes an approach that intrinsically renders this problem irrelevant. Second, there is a logarithmic scaling of the complexity of the design as a function of the required fractional resolution. This scaling feature is a major advantage compared with a standard QHARS device. The latter can typically integrate various hundreds of distinct multicontact Hall bars, whereas a bisection circuit can output many values while only needing a small number of elements to be interconnected. The approach is thought to match, or become comparable to, the present limits of QHR standards.

The design does have some limitations, as pointed out by Momtaz *et al*. They noted that, even though the last bisection stages of the device were controlling a finer portion of the output value, each stage still relied on QHE states crossing a junction barrier, emphasizing the importance of the quality of the junction itself. Another concern of their general design has to do with error propagation caused by imperfections in the device. Their preliminary numerical estimates suggest that such error propagation would be partially fixable since any absolute errors caused by imperfect mixing would not be amplified through the remaining bisection stages and could thus be partitioned by the bisection steps. Nonetheless, this design warrants further study as one way to expand on available quantized resistance outputs.

## 2. Using Arrays to Expand the Parameter Space

Given the difficulty in making *pn*J devices larger than the order 100 μm, which is presently limited by the size of high-quality, single-crystal *h*-BN that can be used as an insulating material for top gating, other approaches to obtaining new resistance outputs must be explored. Recent developments have utilized superconducting materials like NbN and NbTiN to create contact pads compatible with EG-based QHR devices [67-69]. These metals have a high enough critical temperature (about 10 K) and a high enough critical field (about 20 T) that they may be applied to QHR devices and maintain their properties during measurement. These properties have enabled further work, as shown by Kruskopf *et al* [69]. They argue that array technology based on superconducting metals is preferred because other resistance networks based on multiple QHR devices suffer from accumulated resistances at contacts and interconnections. They demonstrated that the application of NbTiN, along with superconducting split contacts, enabled both four terminal and two terminal precision measurements. The split contacts are inspired by a similar approach described by Delahaye [60], and can also be described as multi-series interconnections. And since the device contact resistances now become much smaller than $R_K$, it becomes straightforward that one could access, with high precision, new resistance outputs by using series and parallel connections as fundamental units. The limits of this technology have not yet been determined, as one may expect that, eventually, complex enough arrays would invite an accumulation of resistances at electrical contact pads large enough to disrupt precision measurements. Nonetheless, the most up-to-date literature seem to point at these structures as the next generation of QHR devices.

Another example of array technology that expands on the parameter space comes from Park *et al*. [70], where they sucessfully construct 10 single Hall bars in series with EG on SiC. They operated this device at the ν = 2 plateau and were able to output about 129 kΩ with precision measurements performed using a CCC. While measuring the device at a magnetic field of 6 T and a temperature of 4 K, they were able to achieve an uncertainty of approximately $4 \times 10^{-8}$. Despite only being able to inject a low double-digit electrical current (in μA), their efforts added support to the notion of expanding QHR values with QHARS devices. One difficulty that could arise from making these arrays, especially as they increase in lateral size, is how to make their carrier densities uniform. For that, there are two prime examples of accomplishing this task, with both methods being user-friendly and attaining a long shelf life for the device. The first method involves functionalizing the EG surface with Cr(CO)$_3$ [52], and the second involves a polyer-assisted doping process [53].

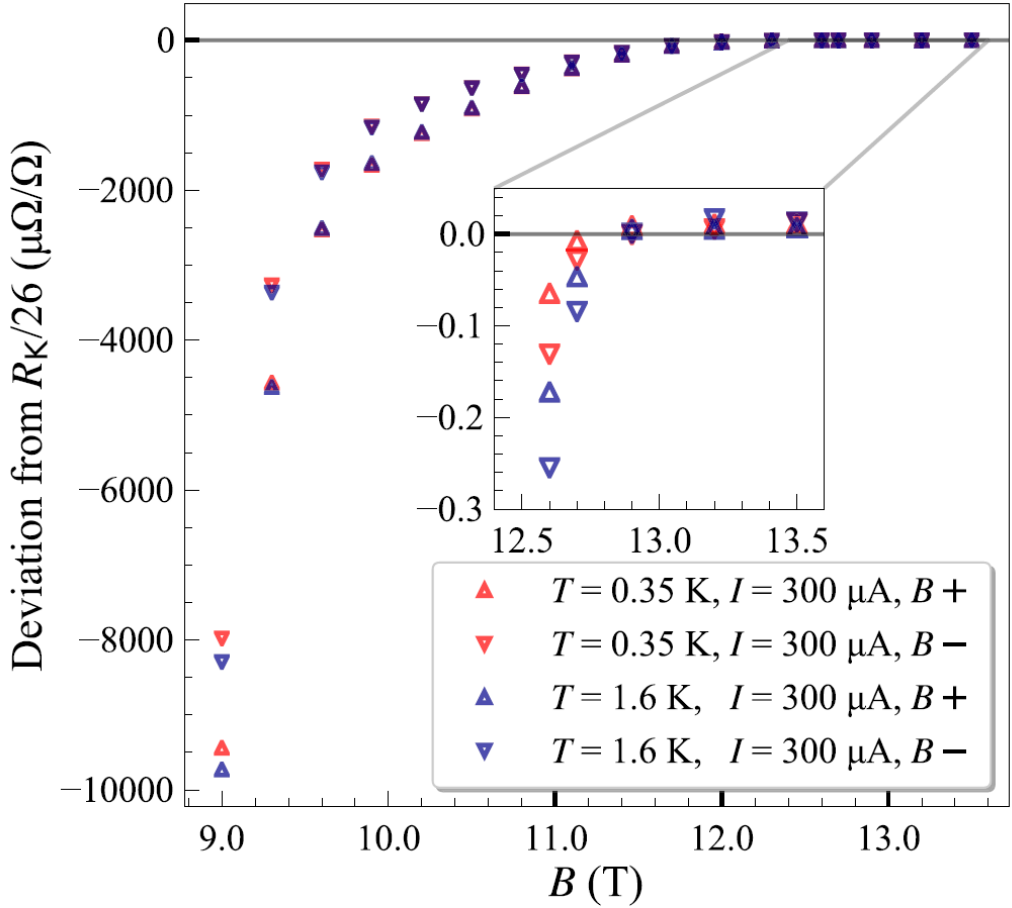

Figure 2. The $v = 2$ plateau was measured at selected values of the magnetic field ($B$) for device 1 at two temperatures (0.35 K and 1.6 K) in a helium-3 cryostat using a room-temperature DCC. The carrier density was determined to be $7 \times 10^{11}$ cm$^{-2}$. The inset shows the data on a magnified scale, where the fully quantized plateau is seen to begin around 12.8 T and 13 T for positive and negative magnetic field polarities, respectively. In general, the resistance values for positive magnetic fields are closer to the quantized value for field values shown in the inset, but at somewhat lower fields the negative magnetic field results are displaced closer to the plateau by increased longitudinal resistance. All expanded uncertainty values given here are for a $2\sigma$ confidence interval. Ref. [61] is an open access article distributed under the terms of the Creative Commons CC BY license, which permits unrestricted use, distribution, and reproduction in any medium.

With the ability to make very large QHARS devices having controllable and uniform carrier densities, researchers were able to construct a 1 kΩ array based on 13 single Hall bar elements in parallel [61]. A CCC was used to measure two 13-element arrays. For 0.3 mA and 1.6 K, the array devices achieved useful quantization above 7 T and 12.5 T. One such array measurement is shown in Fig. 2. In the case of the better of the two arrays, for positive and negative magnetic field polarities, respectively, the deviations from the nominal value of approximately 992.8 Ω were (-0.65 ± 6.32) nΩ/Ω and (-0.25 ± 6.32) nΩ/Ω. As is typical, the CCC ratio uncertainty was below one part in $10^9$, and most of the uncertainty had originated from the 100 Ω artifact resistance standards.

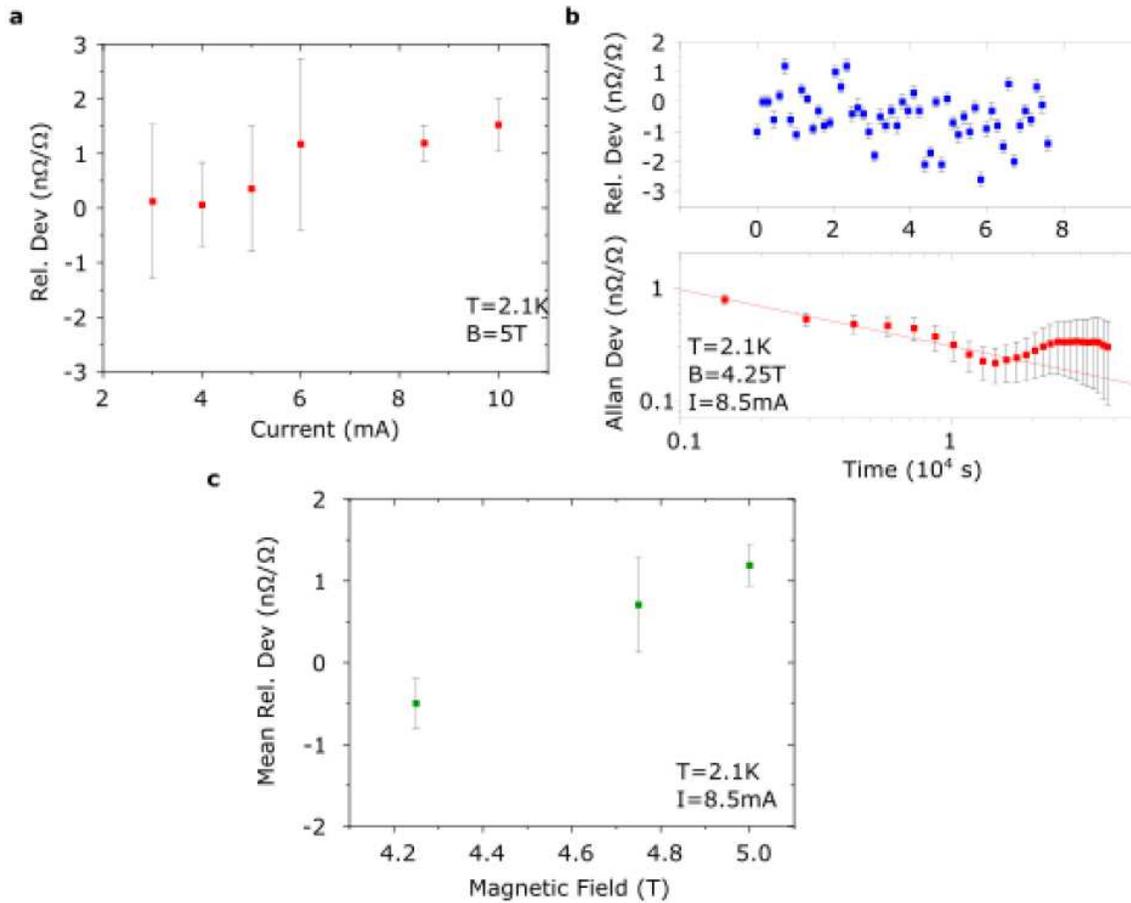

Figure 3. High bias current measurements on arrays. (a) CCC measurements are shown of a direct comparison between subarrays to show that no significant deviation occurs until 8.5 mA. The data consist of the mean of 5 to 10 CCC readings, each of which is 20 min long. Because of this, there is no Allan deviation analysis, and the error bars are thus one standard deviation. (b) Longer precision measurements were taken and also include the Allan deviation. The top graph shows the relative deviation, where each point is a 20 min long CCC reading, with error bars of one standard deviation. The bottom graph shows the corresponding Allan deviation. The standard error is limited to 0.25 nΩ/Ω. (c) The mean relative deviation was calculated from measurements like those in (b). The error bars represent one standard error, taken from the Allan deviation at 104 s. There appears to be a significant deviation at 5 T, which disappears into the uncertainty ($k = 2$) at lower fields. Ref. [68] is an open access article distributed under the terms of the Creative Commons CC BY license, which permits unrestricted use, distribution, and reproduction in any medium.

Recent work by He *et al.* shows quantum Hall measurements performed on a large QHARS device containing 236 individual EG Hall bars [68]. Given the difficulty of verifying the longitudinal resistance as zero for every element, they proposed a direct comparison between two EG-based QHARS devices to verify the accuracy of quantization. The two QHARS devices were compared using high precision measurements, showing no significant deviation of their output resistance within 0.2 nΩ/Ω.

Further measurements are performed to validate the results of the comparison by using a secondary 100 Ω resistance standard. The design of this array is such that two subarrays with 118 parallel elements are connected in series, having a nominal resistance of $h/236e^2$ (about 109.4 Ω), assuming operation at the ν = 2 plateau. The Hall bars were designed to be circular in order to both have a symmetrical design and to be able to packing many devices into a small area. To maximize the injection current, the diameter of the circular element was designed to be 150 μm, which is longer than the equilibration length of QHE edge states at 2 K and 5 T [68]. The contact pads and interconnections were fabricated from NbN given the benefits that superconductivity have been shown to provide to these kinds of samples, and measured at least 120 nm thick and 50 μm wide to support currents on the order of 10 mA at 2 K and 5 T. Additionally, a split contact design was implemented to minimize the contact resistance, much like other previously reported work [67, 69]. Within the next few years, given the increasing complexity of EG-based QHARS devices, even larger and more versatile arrays are expected to make available an abundance of new quantized resistance values.

## 3. From DC to AC to the Quantum Ampere

As the *quantum* SI continues to expand in its permeation of our everyday lives, new research directions are plentiful and of interest to the electrical metrology community. The QHE will continue to be our foundation for disseminating the ohm in the DC realm. The vast improvements in EG-based QHR technology are now beginning to inspire efforts for developing more sophisticated devices suitable for AC resistance standards. This subfield of electrical metrology focuses on the calibration of impedance and is conventionally obtained from systems like a calculable capacitor [71, 72]. Such a system has allowed for the calibration of capacitors, inductors, or AC resistors, which is essentially a measurement of complex ratios of impedance. Historically, the building and operation of this kind of system have been challenging because of unavoidable fringe electric fields and imperfections in capacitor electrode construction.

The next step for improving AC standards may be to introduce EG-based QHR devices, as Kruskopf *et al*. have done recently [73]. They used a conventional EG-based device to analyze the experienced losses and to determine the characteristic capacitances. The straightforward way to achieve this in the device was to use part of the double shield as an electrode, as shown in Fig. 4 (a). Fig. 4 (b) displays the set of magnetocapacitance measurements corresponding to a configuration of the capacitance ($Cx$) between the active electrode (left side of (a)) and the EG Hall bar between points A and B in Fig. 4 (a). $Cx$ was compared with a 1422-CD variable precision reference capacitor using a simple configuration used for other traditional measurements [73]. On the right side of Fig. 4 (a) shows the passive electrode on right side, which was shorted to the EG, therefore not contributing anything to the measurement of $Cx$. In that same figure, one can see the puddles of differing color meant to indicate, in a very abstract way, the regions that are compressible or incompressible.

Kruskopf *et al*. show the voltage and frequency dependencies of the magnetocapacitance in Fig. 4 (b) and (c), respectively, along with the associated dissipation factor for an example device. By appropriately modelling the compressible and incompressible states, the observed dissipation factors may be explained rather well. At low magnetic field values, the 2D electron system in the EG not quantized and is thus accurately representable as a semi-metal with dominant compressible states. However, when the magnetic field is increased, $Cx$ starts to decrease around 4 T and does so by nearly 3 fF in both cases as 12 T is approached. This phenomenon may be due to the increase in incompressible regions that form, which are themselves transparent to electric fields. Furthermore, the dissipation factor is observed to first increase, peaking during the transition region as the resistance plateau in EG forms. As the quantization accuracy improves at higher magnetic fields, the observed dissipative losses decrease again to about tan($\delta$) = 0.0003. Overall, these are but a few magnetocapacitance measurements that demonstrate the viability of our efforts to better understand the physical phenomena driving the observations made in the QHE regime while under AC conditions. This pursuit of AC QHR standards is hopefully bound to lead to advancements in how we realize units such as the farad and henry by using fundamental constants instead of dimensional measurements. Similar work has been done to directly access the units of capacitance and inductance with high precision [74].

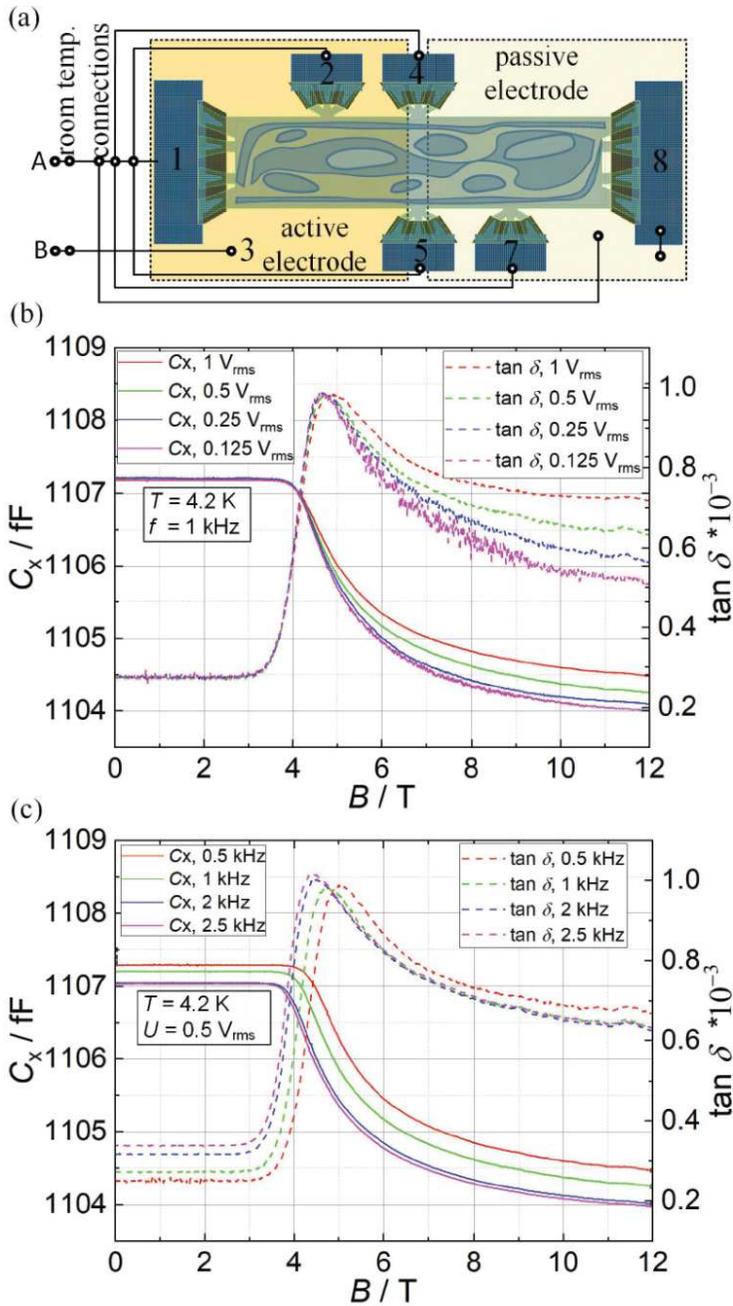

Figure 4. Magnetocapacitance measurement data of an example device are shown (taken at 4.2 K). (a) An illustration captures the various elements of the used magnetocapacitance measurement configuration. As one electrode gets shorted to the labelled passive electrode (pin 8), the other electrode is then used to characterize C and tan($\delta$). Compressible and incompressible states of the 2D electron system are represented as different colors within the EG Hall bar. (b) The capacitance is plotted as a function of magnetic field for a set of different voltages. (c) A similar plot to (b) is shown for a set of different frequencies. The dissipation factor (tan($\delta$)) is also plotted representing the losses between active electrode and the EG Hall bar device. Ref. [73] is an open access article distributed under the terms of the Creative Commons CC BY license, which permits unrestricted use, distribution, and reproduction in any medium.

In addition to expanding the influence of the QHE to AC electrical metrology, one may also expand into the realms of electrical current and mass metrology by utilizing EG-based QHARS devices. The realization of the quantum ampere lacks accurate traceability to within $10^{-8}$ despite the various efforts that exist for developing a current source using single-electron tunneling devices [75-77]. The alternative route one may take to realizing the quantum ampere is by building a circuit that effectively combines the QHE and a programmable Josephson junction voltage standard. This combination has been recently assembled to attain a programmable quantum current generator, which may disseminate the ampere at the milliampere range and above [78]. The work reported by Brun-Picard *et al*. demonstrates this construction with a superconducting cryogenic amplifier and to measurement uncertainties of $10^{-8}$. Their quantum current source, which is housed as two separate cryogenic systems, can deliver accurate currents down to the microampere range. Considering the orders of magnitude involves, this work renders the programmable quantum current generator a pronounced supplement to electron-pumping mechanisms.

It has become evident throughout the last few decades that the quantum Hall effect, as exhibited by our modern 2D systems both with and without magnetic fields, has the marvelous potential to unify the components of Ohm's law. That is, the QHE can bring together all three electrical quantities provided that the magnetic field requirement is small or irrelevant to the apparatus in use. Developing and deploying a system with several traceability capabilities will undoubtedly improve the status of electrical metrology worldwide. Throughout all the coming advancements [79-84], it will be important to remember that these milestones should keep us motivated to continue learning how to better enrich society with the quantum Hall effect:

> "It is characteristic of fundamental discoveries, of great achievements of intellect,
> that they retain an undiminished power upon the imagination of the thinker."
> – Nikola Tesla, 1891, New York City, New York

**Acknowledgements**

The authors wish to acknowledge S. Mhatre, A. Levy, G. Fitzpatrick, and E. Benck for their efforts and assistance during the internal review process at NIST. Commercial equipment, instruments, and materials are identified in this paper in order to specify the experimental procedure adequately. Such identification is not intended to imply recommendation or endorsement by the National Institute of Standards and Technology or the United States government, nor is it intended to imply that the materials or equipment identified are necessarily the best available for the purpose.

**References**

[1] v Klitzing K, Ebert G. Application of the quantum Hall effect in metrology. Metrologia. 1985;21(1):11.
[2] Von Klitzing K, Dorda G and Pepper M 1980 *Phys. Rev. Lett.* **45** 494
[3] Hill HM, et al. 2019 *Phys. Rev. B.* **99** 174110
[4] Witt TJ 1998 *Rev. Sci. Instrum.* **69** 2823-43
[5] Hartland A, Jones K, Williams J M, Gallagher B L and Galloway T 1991 *Phys. Rev. Lett.* **66** 969-73
[6] Cage ME, Dziuba RF, Field BF 1985 *IEEE Trans. Instrum. Meas.* **2,** 301-3.
[7] Hartland A 1992 *Metrologia* **29** 175
[8] Tsui D C and Gossard A C 1981 *Appl. Phys. Lett.* **38** 550
[9] Van Der Wel W, Harmans KJ, Kaarls R, Mooij JE 1985 *IEEE Trans. Instrum. Meas*. **2** 314-6
[10] Hartland A, Jones R G, Kibble B P and Legg D J 1987 *IEEE Trans. Instrum. Meas.* **IM-36** 208
[11] Bliek L, Braun E and Melchert F 1983 *Metrologia* **19**, 83
[12] Hartland A, Davis GJ and Wood DR 1985 *IEEE Trans. Instrum. Meas.* **IM-34** 309
[13] Delahaye F, Dominguez D, Alexandre F, André J P, Hirtz J P and Razeghi M 1986 *Metrologia* **22** 103-10
[14] Taylor BN 1990 *IEEE Trans. Instrum. Meas.* **39** 2-5
[15] Jeckelmann B, Jeanneret B and Inglis D. 1997 *Phys. Rev. B* **55** 13124
[16] Williams JM IET 2011 *Sci. Meas. Technol.* **5** 211-24
[17] Hamon BV 1954 *J. Sci. Instrum.* **31** 450-453
[18] Small GW, Ricketts BW, and Coogan PC 1989 *IEEE Trans. Instrum. Meas.* **38** 245
[19] Cage ME, Dziuba RF, Elmquist RE, Field BF, Jones GR, Olsen PT, Phillips WD, Shields JQ, Steiner RL, Taylor BN, and Williams ER 1989 *IEEE Trans. Instrum. Meas.* **38** 284
[20] Shields JQ and Dziuba RF 1989 *IEEE Trans. Instrum. Meas.* **38** 249
[21] Delahaye F and Jeckelmann B 2003 *Metrologia* **40** 217
[22] Jeckelmann B and Jeanneret B 2001 *Rep. Prog. Phys.* **64** 1603
[23] Oe T, Matsuhiro K, Itatani T, Gorwadkar S, Kiryu S, Kaneko NH 2013 *IEEE Trans. Instrum. Meas.* **62** 1755-9
[24] Poirier W, Bounouh A, Piquemal F and André JP 2004 *Metrologia* **41** 285
[25] Ortolano M, Abrate M and Callegaro L 2014 *Metrologia* **52** 31
[26] Konemann J, Ahlers FJ, Pesel E, Pierz K and Schumacher HW 2011 *IEEE Trans. Instrum. Meas.* **60** 2512-6
[27] Ahlers FJ, Jeanneret B, Overney F, Schurr J, Wood BM 2009 *Metrologia* **46** R1
[28] Cabiati F, Callegaro L, Cassiago C, D'Elia V, Reedtz GM. 1999 *IEEE Trans. Instrum. Meas*. **48** 314-8
[29] Bykov AA, Zhang JQ, Vitkalov S, Kalagin AK, Bakarov AK 2005 *Phys. Rev. B*. **72** 245307
[30] Hartland A, Kibble BP, Rodgers PJ, Bohacek J. 1995 *IEEE Trans. Instrum. Meas.* **44** 245-8
[31] Wood HM, Inglis AD, Côté M 1997 *IEEE Trans. Instrum. Meas* **46** 269-72
[32] Zhang Y, Tan YW, Stormer HL and Kim P 2005 Nature **438** 201
[33] Novoselov KS, Jiang Z, Zhang Y, Morozov SV, Stormer HL, Zeitler U, Maan JC, Boebinger GS, Kim P, Geim AK 2007 Science **315** 1379
[34] Novoselov KS, Geim AK, Morozov S, Jiang D, Katsnelson M, Grigorieva I, Dubonos S, Firsov AA 2005 Nature **438** 197
[35] De Heer WA, Berger C, Wu X, First PN, Conrad EH, Li X, Li T, Sprinkle M, Hass J, Sadowski ML, Potemski M 2007 *Solid State Commun*. **143** 92-100
[36] Jabakhanji B, Michon A, Consejo C, Desrat W, Portail M, Tiberj A, Paillet M, Zahab A, Cheynis F, Lafont F, Schopfer F. 2014 *Phys. Rev. B* **89** 085422
[37] Janssen TJ, Williams JM, Fletcher NE, Goebel R, Tzalenchuk A, Yakimova R, Lara-Avila S, Kubatkin S, Fal'ko VI 2012 *Metrologia* **49** 294
[38] Giesbers AJ, Rietveld G, Houtzager E, Zeitler U, Yang R, Novoselov KS, Geim AK, Maan JC 2008 *Appl. Phys. Lett.* **93** 222109–12
[39] Tzalenchuk A Lara-Avila S, Kalaboukhov A, Paolillo S, Syväjärvi M, Yakimova R, Kazakova O, Janssen TJ, Fal'Ko V, Kubatkin S 2010 *Nat. Nanotechnol.* **5** 186–9
[40] Lafont F, Ribeiro-Palau R, Kazazis D, Michon A, Couturaud O, Consejo C, Chassagne T, Zielinski M, Portail M, Jouault B, Schopfer F 2015 *Nat. Commun.* **6** 6806
[41] Janssen T J B M, Tzalenchuk A, Yakimova R, Kubatkin S, Lara-Avila S, Kopylov S, Fal'ko VI. 2011 *Phys. Rev. B* **83** 233402–6
[42] Oe T, Rigosi AF, Kruskopf M, Wu BY, Lee HY, Yang Y, Elmquist RE, Kaneko N, Jarrett DG 2019 *IEEE Trans. Instrum. Meas.* 2019 **69** 3103-3108
[43] Woszczyna M, Friedemann M, Götz M, Pesel E, Pierz K, Weimann T, Ahlers FJ 2012 *Appl. Phys. Lett.* **100** 164106
[44] Satrapinski A, Novikov S, Lebedeva N 2013 *Appl. Phys. Lett.* **103** 173509
[45] Rigosi AF, Panna AR, Payagala SU, Kruskopf M, Kraft ME, Jones GR, Wu BY, Lee HY, Yang Y, Hu J, Jarrett DG, Newell DB, and Elmquist RE 2019 *IEEE Trans. Instrum. Meas*. **68**, 1870-1878.
[46] Lara-Avila S, Moth-Poulsen K, Yakimova R, Bjørnholm T, Fal'ko V, Tzalenchuk A, Kubatkin S 2011 *Adv. Mater.* **23** 878-82
[47] Rigosi AF, Liu CI, Wu BY, Lee HY, Kruskopf M, Yang Y, Hill HM, Hu J, Bittle EG, Obrzut J, Walker AR 2018 Microelectron. Eng. **194** 51-5
[48] Riedl C, Coletti C, Starke U 2010 *J. Phys. D* **43** 374009

[49] Rigosi AF, Hill HM, Glavin NR, Pookpanratana SJ, Yang Y, Boosalis AG, Hu J, Rice A, Allerman AA, Nguyen NV, Hacker CA, Elmquist RE, Newell DB 2017 *2D Mater*. **5** 011011
[50] Janssen TJ, Rozhko S, Antonov I, Tzalenchuk A, Williams JM, Melhem Z, He H, Lara-Avila S, Kubatkin S, Yakimova R 2015 *2D Mater.* **2** 035015
[51] Kruskopf M, Pakdehi DM, Pierz K, Wundrack S, Stosch R, Dziomba T, Götz M, Baringhaus J, Aprojanz J, Tegenkamp C, Lidzba J. 2016 *2D Mater*. **3** 041002
[52] Hill HM, Rigosi AF, Chowdhury S, Yang Y, Nguyen NV, Tavazza F, Elmquist RE, Newell DB, Walker AR. 2017 *Phys. Rev. B* **96** 195437
[53] Rigosi AF, Elmquist RE 2019 *Semicond. Sci.Ttechnol*. 2019 **34** 093004
[54] MacMartin MP, Kusters NL 1996 *IEEE Trans. Instrum. Meas.* **15** 212-20
[55] Drung D, Götz M, Pesel E, Storm JH, Aßmann C, Peters M, Schurig T 2009 *Supercond. Sci. Technol.* **22** 114004
[56] Sullivan DB and Dziuba RF 1974 *Rev. Sci. Instrum*. **45** 517
[57] Grohmann K, Hahlbohm HD, Lübbig H, and Ramin H 1974 *Cryogenics* **14** 499
[58] Williams J M, Janssen T J B M, Rietveld G and Houtzager E 2010 *Metrologia* **47** 167–74
[59] Zhang N 2006 *Metrologia* **43** S276-S281
[60] Delahaye F 1993 *J. Appl. Phys*. **73** 7914-20
[61] Panna AR, Hu IF, Kruskopf M, Patel DK, Jarrett DG, Liu CI, Payagala SU, Saha D, Rigosi AF, Newell DB, Liang CT 2021 *Phys. Rev. B* **103** 075408
[62] Hu J, Rigosi AF, Kruskopf M, Yang Y, Wu BY, Tian J, Panna AR, Lee HY, Payagala SU, Jones GR, Kraft ME, Jarrett DG, Watanabe K, Takashi T, Elmquist RE, Newell DB 2018 *Sci. Rep*. **8** 15018
[63] Woszczyna M, Friedemann M, Dziomba T, Weimann T, Ahlers FJ 2011 *Appl. Phys. Lett*. **99** 022112
[64] Hu J, Rigosi AF, Lee JU, Lee HY, Yang Y, Liu CI, Elmquist RE, Newell DB 2018 *Phys. Rev B* **98** 045412
[65] Rigosi AF, Patel DK, Marzano M, Kruskopf M, Hill HM, Jin H, Hu J, Hight Walker AR, Ortolano M, Callegaro L, Liang CT, Newell DB 2019 *Carbon* **154** 230-237
[66] Momtaz ZS, Heun S, Biasiol G, Roddaro S 2020 *Phys. Rev. Appl.* **14** 024059
[67] Kruskopf M, Rigosi AF, Panna AR, Marzano M, Patel DK, Jin H, Newell DB, Elmquist RE 2019 *Metrologia* **56** 065002
[68] He H, Cedergren K, Shetty N, Lara-Avila S, Kubatkin S, Bergsten T, Eklund G 2021 *arXiv* preprint arXiv:2111.08280
[69] Kruskopf M, Rigosi AF, Panna AR, Patel DK, Jin H, Marzano M, Newell DB, Elmquist RE 2019 *IEEE Trans. Electron Devices* **66** 3973-3977
[70] Park J, Kim WS, Chae DH 2020 *Appl. Phys. Lett.* **116** 093102
[71] Clothier WK 1965 *Metrologia* **1** 36
[72] Cutkosky RD 1974 IEEE Trans. Instrum. Meas. **23** 305-9
[73] Kruskopf M, Bauer S, Pimsut Y, Chatterjee A, Patel DK, Rigosi AF, Elmquist RE, Pierz K, Pesel E, Götz M, Schurr J 2021 *IEEE Trans. Electron Devices* **68** 3672-3677
[74] Lüönd F, Kalmbach CC, Overney F, Schurr J, Jeanneret B, Müller A, Kruskopf M, Pierz K, Ahlers F 2017 *IEEE Trans. Instrum. Meas.* **66** 1459-66
[75] Giblin SP, Kataoka M, Fletcher JD, See P, Janssen TJ, Griffiths JP, Jones GA, Farrer I, Ritchie DA 2012 *Nat. Commun.* 2012 **3** 1-6
[76] Pekola JP, Saira OP, Maisi VF, Kemppinen A, Möttönen M, Pashkin YA, Averin DV 2013 *Rev. Mod. Phys.* **85** 1421
[77] Koppinen PJ, Stewart MD, Zimmerman NM 2012 *IEEE Trans. Electron Devices* **60** 78-83
[78] Brun-Picard J, Djordjevic S, Leprat D, Schopfer F, Poirier W 2016 *Phys. Rev. X* **6** 041051
[79] Rigosi AF, Marzano M, Levy A, Hill HM, Patel DK, Kruskopf M, Jin H, Elmquist RE, Newell DB 2020 *Phys. B: Condens. Matter* **582** 411971
[80] Waldmann D, Jobst J, Speck F, Seyller T, Krieger M, Weber HB 2011 *Nat. Mater.* **10** 357-60
[81] Fox EJ, Rosen IT, Yang Y, Jones GR, Elmquist RE, Kou X, Pan L, Wang KL, and Goldhaber-Gordon D 2018 *Phys. Rev. B* **98** 075145
[82] Götz M, Fijalkowski KM, Pesel E, Hartl M, Schreyeck S, Winnerlein M, Grauer S, Scherer H, Brunner K, Gould C, Ahlers FJ 2018 *Appl. Phys. Lett.* **112** 072102
[83] Fijalkowski KM, Liu N, Mandal P, Schreyeck S, Brunner K, Gould C, Molenkamp LW 2021 *Nat. Commun.* 2021 **12** 1-7
[84] Okazaki Y, Oe T, Kawamura M, Yoshimi R, Nakamura S, Takada S, Mogi M, Takahashi KS, Tsukazaki A, Kawasaki M, Tokura Y 2021 *Nat. Phys.* 13:1-5